\documentstyle[11pt]{article}

\begin{document}
\oddsidemargin .1in
\evensidemargin 0 true pt
\topmargin -.4in


\def\ra{{\rightarrow}}
\def\a{{\alpha}}
\def\b{{\beta}}
\def\l{{\lambda}}
\def\eps{{\epsilon}}
\def\T{{\Theta}}
\def\t{{\theta}}
\def\co{{\cal O}}
\def\car{{\cal R}}
\def\caf{{\cal F}}
\def\cs{{\Theta_S}}
\def\pr{{\partial}}
\def\tri{{\triangle}}
\def\na{{\nabla }}
\def\S{{\Sigma}}
\def\s{{\sigma}}
\def\sp{\vspace{.15in}}
\def\hs{\hspace{.25in}}

\newcommand{\be}{\begin{equation}} \newcommand{\ee}{\end{equation}}
\newcommand{\bea}{\begin{eqnarray}}\newcommand{\eea}
{\end{eqnarray}}


\begin{titlepage}
\topmargin= -.2in
\textheight 9in
\textwidth 6.5in
\begin{center}
\baselineskip= 17 truept

\vspace{.3in}
\centerline{\Large\bf Emergent gravity/Non-linear $U(1)$ gauge theory correspondence} 


\vspace{.6in}

\noindent
{{{\large Supriya Kar}\footnote{skkar@physics.du.ac.in }},
{{\large K. Priyabrat Pandey}\footnote{kppandey@physics.du.ac.in }},
{{\large Abhishek K. Singh}\footnote{aksingh@physics.du.ac.in }} {{\large and Sunita Singh}\footnote{sunita@physics.du.ac.in }}}

\vspace{.2in}

\noindent
{\large Department of Physics \& Astrophysics\\
{\large University of Delhi, New Delhi 110 007, India}}

\vspace{.2in}

{\today}
\thispagestyle{empty}

\vspace{.6in}
\begin{abstract}

\vspace{.2in}
Kaluza-Klein gravity is revisited, with renewed interest, in a type IIB string theory on $S^1\times K3$. The irreducible curvature tensors are 
worked out in the, $T$-dual, emergent gravity in $4D$ to yield a non-linear $U(1)$ gauge theory. Interestingly, the $T$-duality 
may be seen to describe an open/closed string duality at a self-dual string coupling. The obtained deformation in  $AdS_5$ black hole is analyzed 
to introduce the notion of temperature in the emergent gravity underlying the recent idea of entropic force.

\end{abstract}
\end{center}

\vspace{.2in}

\baselineskip= 14 truept

\vspace{1in}



\end{titlepage}

\baselineskip= 18 truept

\section{Introduction}
The intrinsic notion of geometry, governing space-time curvature, is manifested beautifully in the success of the General Theory of Relativity (GTR). Importantly, the Einstein's field equations confirm the presence of non-linear matter and establishes a precise relation between the
matter and geometry in GTR. With the help of an extra dimension to Einstein's theory, the matter/geometry relation may be explored to view as a gauge/gravity duality or in particular $AdS_5$-gravity/gauge theory duality established in literatures \cite{maldacena,witten,li}.

\vspace{.15in}
Very recently, it has been conjectured that gravity emerges at a macroscopic level due to the entropic force described by the
the matter field, which in turn results in a deformed geometry \cite{verlinde-10}. It is plausible that the deformation may be described by $D$-branes \cite{kpss} contained within higher dimensional branes in string theory \cite{douglas,witten-bound,polyakov}. Similar to AdS/CFT, the emergent gravity is believed to be projected along a preferred holographic direction in space, which possibly governs a cut-off scale on the boundary. Since a flat space is not associated with any holographic direction, it is promising to begin with Newtonian potential to describe the emergent gravity phenomenon as a non-relativistic limit of GTR. As a result, the holographic direction may be seen to be governed by a gradient of the scalar potential, which in turn describes an accelerated frame and establishes the notion of temperature \cite{unruh}. 

\vspace{.15in}
In the article, we attempt to focus on the notion of emergent gravity in $4D$ due to the two-form underlying a string theory in presence of $D_3$-branes. In section 2.1, we elaborate on the idea of emergent gravity underlying ``Kaluza-Klein'' theory. In section 2.2, we re-investigate a deformed $AdS_5$ black hole geometry in the emergent gravity to emphasize the notion of temperature in the formalism. We obtain the dual actions in $4D$ by taking into account the irreducible gauge and gravity curvature tensors in section 3. We conclude the article in section 4.

\section{Emergent gravity in $5D$ due to a two-form}

\subsection{``Kaluza-Klein'' gravity}
We begin with Kaluza-Klein gravity in presence of a cosmological constant $\Lambda\neq 0$.
The action is given by
\be
S= {1\over{G_N}}\int d^5x \ {\sqrt{-G}}\ \left (R^{(5)}-\Lambda\right )\ ,\label{kk-1}
\ee
where $G_N$ denotes the (appropriate) Newton's constant.
We revisit the Einstein's geometric notion in GTR, by incorporating a covariantly constant dynamical two-form $B_{mn}$ along with a non-dynamical metric tensor $g_{mn}$ in the Kaluza-Klein gravity theory. Interestingly, the typical notion of gravity leading to nontrivial geometry may be seen to be prevailed with the source of two-form. In fact, the dynamical metric in Kaluza-Klein theory may alternately be obtained from a two-form using
a relation:
\be
G_{mn} = g_{mn} - \left ( B_{mp}g^{pq}B_{qn}\right )\ .\label{kk-2}
\ee
A priori, the ``Kaluza-Klein'' gravity due to the two-form source may be given by
\be
S={1\over{G_N}}\int d^5x \ {\sqrt{-g}} \ K^{(5)} \ .\label{kk-3}
\ee
The cosmological constant becomes redundant in the two-form source action. However in the emergent gravity (\ref{kk-2}) description, $\Lambda\neq 0$ becomes significant. The irreducible scalar curvature $K$ is obtained from the 4th order mixed curvature tensor. It is given by 
\be
{K_{mnp}}^q= \partial_n{\mathbf\Gamma}^q_{mp} -\partial_m{\mathbf\Gamma}^q_{np} + {\mathbf\Gamma}^s_{mp}{\mathbf\Gamma}^q_{ns}-
{\mathbf\Gamma}^s_{np}{\mathbf\Gamma}^q_{ms}\ .\label{kk-4}
\ee
The new connections in the formalism may be expressed as:
\bea
&&{\mathbf\Gamma}_{mnp} = {\mathbf\Gamma}^q_{mn}B_{qp} + {\rm cyclic}\; ({\rm in}\ m, n, p)\nonumber\\
{\rm and} &&{\mathbf\Gamma}^q_{mn}= -{1\over2} g^{qs}\left ( \partial_mB_{sn} + {\rm cyclic} \right ) \ .
\label{kk-5}
\eea
The covariant derivative is uniquely fixed by ${\cal D}_pg_{mn}=0$ and ${\cal H}_{mnp}=\left ({\cal D}_mB_{np} + {\rm cyclic}\right )=0$ in the formalism. It is important to note that the generic curvature tensor ${K_{mnp}}^q$ modifies the Riemann-Christoffel curvature ${R_{mnp}}^q$ in Kaluza-Klein gravity. However for a constant torsion $T^s_{mn}=-2{\mathbf\Gamma}^s_{mn}$, the geometric curvature ${K_{mnp}}^q$ precisely reduces to
${R_{mnp}}^q$ and $\Lambda\neq 0$. In other words, the Riemannian notion of geometry leading to a typical Kaluza-Klein gravity is restored for a constant torsion. As a result, a constant torsion geometry in the formalism is indistinguishable from the dynamical metric geometry. The field equations may be given by
\be
\partial_p\Gamma^{pmn} - {1\over2} \Big ( g^{ab}\partial_{p}\ g_{ab}\Big ) \Gamma^{pmn}=0
\ .\label{kk-6}
\ee
\subsection{Notion of temperature from deformed $AdS_5$}
We consider the $B$-field ansatz in obtained recently in ref.\cite{kpss}. They are given by:
\be
B_{0\psi}= B_{r\psi}=r_0\ ,\qquad B_{\theta\psi}= q (\cot\theta\sin^2\psi)\qquad {\rm and}\quad B_{\psi\phi}= q
(\cos\theta\sin^2\psi)\ ,\label{kk-7}
\ee
where $0<\theta\le\pi$, $0<\psi\le\pi$, $0\le\phi\le 2\pi$ and ($r_0$ and $q$) are are arbitrary constants. The non-vanishing components of torsion become
\be
{\mathbf\Gamma}_{\theta\psi\phi}= {{q}\over{2}}(\sin\theta\sin^2\psi)\ {\rm and}\quad 
{\mathbf\Gamma}_{\theta 0\phi}= {\mathbf\Gamma}_{\theta r\phi} = {{(qr_0)}\over{2r^2}}(\sin\theta\sin^2\psi)\ .\label{kk-8}
\ee
The geodesic describing an $AdS_5$ black hole may be obtained from the two-form using the relation (\ref{kk-2}) around
an $AdS$ vacua. The $AdS_5$ geometry is given by 
\bea
&&ds^2=-\left (1+{{r^2}\over{b^2}}-{{r_0^2}\over{r^2}}\right ) dt^2 + \left ( 1 + {{r^2}\over{b^2}} -{{r_0^2}\over{r^2}}\right )^{-1} dr^2
+r^2 \left ( 1+ {{2f^2}\over{r^4}}\right ) d\psi^2\qquad\qquad\qquad\qquad {}\nonumber\\
&&\qquad\qquad\qquad\qquad\qquad\qquad\qquad\qquad\qquad\quad + r^2\sin^2\psi \left ( 1 + {{f^2}\over{r^4}}\right ) \left ( d\theta^2 + \sin^2\theta\ d\phi^2\right )\ .\label{kk-9}
\eea
where $f= q(\cot\theta\sin\psi)$ and $b$ denotes the $AdS$ radius. The deformed $S^3$ geometry in $AdS_5$ black hole is a new feature in the formalism. Importantly, the deformation to the $AdS_5$ Schwarzschild black hole \cite{sen,horowitz-polchinski} in a typical Kaluza-Klein gravity is essentially due to the two-form in the formalism.
In other words, the broken $S^3$-symmetry is restored for $f=0$, $i.e.$ for a constant torsion and the $AdS_5$ geometry (\ref{kk-9}) reduces to the $AdS_2\times S^3$ Schwarzschild black hole geometry. In fact, the $S^3$ deformation, by an $S^2$, is independent of the torsion and is solely a characteristic of the non-vanishing two-form. Since a dynamical two-form is dual to a scalar potential, the notion of temperature ($T$) may be seen to be governed by the deformation geometry in the formalism. For a constant torsion, the curvature is described by the Riemannian geometry and the deformation vanishes leading to $T=0$. It is thought provoking to believe that the association of temperature is possibly defined naturally in any formalism defined with a non-vanishing two-form. For instance, the deformed black hole geometry obtained on a non-commutative $D_3$-brane \cite{karm,kar}, is a consequence of the non-commutative parameter and has been argued to be the origin of non-zero temperature \cite{kar-panda}.

\section{Dual geometries in 4D}
The Kaluza-Klein compactification of the fifth dimension in the action (\ref{kk-1}) is worked out, a priori, to obtain the gravity and $U(1)$ gauge curvatures. The irreducible curvatures, obtained by Kaluza-Klein compactification of action (\ref{kk-3}), give rise to
\be
S= {1\over{G_N}}\int\ d^4x {\sqrt{-g}}\ \left ( K^{(4)} - {{G_N}\over4}\ F^2\right )\ ,\label{kk-10}
\ee
where the geometric scalar curvature $K^{(4)}$ is precisely governed by the torsion in the formalism. On the other hand, the torsion may be thought of as higher order derivative corrections to the electromagnetic field strength in a dual channel \cite{kpss}. Alternately, one may obtain an $U(1)$ gauge invariant field strength by incorporating the torsion. The non-linear $U(1)$ field strength becomes
\be
{\cal F}_{\mu\nu} = F_{\mu\nu} + T^{\lambda}_{\mu\nu}A_{\lambda}\ .\label{kk-11}
\ee
The irreducible curvatures in the action (\ref{kk-10}) may be re-expressed in terms of non-linear gauge curvature in a dual channel. 
The action takes a simple form 
\be
S= -{1\over4}\int\ d^4x {\sqrt{-g}}\ {\cal F}^2\ .\label{kk-12}
\ee
The non-linear $U(1)$ gauge theory in $4D$ dual to ``Kaluza-Klein'' gravity is remarkable. Firstly, it re-assures Einstein's assertion 
that the space-time curvature is completely governed by the non-linear matter field. Secondly, the underlying duality is accompanied with an 
a prior surprise due to its amazing correspondence between ``Kaluza-Klein'' gravity and the non-linear $U(1)$ gauge theory in $4D$. Interestingly, the gravity/gauge theory correspondence is along the line of holographic idea. 

\vspace{.15in}
On the other hand, the apparent duality between the ``Kaluza-Klein'' gravity and gauge theory may be better understood in a closed string theory. For instance, we consider a D=10 type IIB string on $S^1\times K3$ obtained by one of the author in a collaboration \cite{kar-maharana-panda}. 
The NS-NS sector in $5D$ may be worked out for a consistent truncation to yield 
``Kaluza-Klein'' gravity (\ref{kk-3}). In particular, the geometric scalar curvature in the formalism may be identified with the gauge curvature in string theory, $i.e.\ K \equiv -(dB)^2$, in a dual channel with a covariantly constant dynamical two-form. In fact, the obtained $AdS_5$ geometry (\ref{kk-9}) may be viewed as an underlying $AdS_5\times S^5$ geometry in type IIB string theory. This in turn, corresponds to the open string boundary gauge theory on a $D_3$-brane \cite{maldacena}. In other words, the non-linear $U(1)$ gauge theory (\ref{kk-12}) may be identified with 
the $D_3$-brane Born-Infeld dynamics in string theory \cite{kpss}. As a result, the gravity/gauge theory correspondence may be viewed as closed/open string duality on the one hand and the $T$-duality ($R\rightarrow {\alpha'/R}$) in string theory on the other hand. The identification of gravity/gauge theory correspondence under two distinct ($T$- and $S$-) dualities in string theory is unique to the emergent gravity description due to the $B$-field source. This is essentially due to the fact that the underlying $S$-duality is defined at the self-dual string coupling, $i.e.\ g_s=1$, and the $T$-dual fifth dimension in closed string theory is transverse to the $D_3$-braneworld \cite{karm}.

\section{Concluding remarks}
The emergent Kaluza-Klein gravity due to the source of $B$-field is worked out in a type IIB string theory on $S^1\times K3$. Interestingly, the $T$-duality was identified with the $S$-duality at a self-dual string coupling in the formalism. The non-linear $U(1)$ gauge theory obtained in the dual channel was identified with the Born-Infeld dynamics of $D_3$-brane. In addition to the open/closed string duality, the formalism allows one to obtain a dual $D_3$-brane description in $4D$ itself. On the one hand, the electro-magnetic field may be seen to govern the $D_3$-brane dynamics underlying a fundamental open string theory (\ref{kk-10}). On the other hand, a non-linear electro-magnetic field governs a dual $D_3$-brane dynamics underlying a NCOS theory (\ref{kk-12}). Interestingly, the $D_3$-brane and its dual are manifestations of two distinct $U(1)$ gauge symmetries established in literature \cite{seiberg-witten}.

\vspace{.15in}
\noindent
The deformed $AdS_5$ black hole obtained in the emergent gravity description was re-investigated for the notion of temperature. It was shown that a two-form is solely responsible for the deformation geometry which in turn gives rise to a non-zero temperature in the formalism. 
Our analysis may be seen to be in conformity with the recent idea of entropic force underlying an emergent gravity \cite{verlinde-10}. We observe that the Einstein's geometric description of gravity is potentially a powerful tool and may help one to explore new physics in various other \cite{gubser,hartnoll}, a priori, unrelated branches in science.

\vfil\eject
\noindent
{\large\bf Acknowledgments}

\noindent
The work of S.K. is partly supported by a research project under D.S.T, Govt.of India. AKS acknowledges C.S.I.R, New Delhi for a partial support.

\sp

\def\anp{Ann. of Phys.}
\def\prl{Phys.Rev.Lett.}
\def\prd#1{{Phys.Rev.}{\bf D#1}}
\def\jhep{JHEP}
\def\cqg#1{{Class.\& Quant.Grav.}}
\def\plb#1{{Phys. Lett.} {\bf B#1}}
\def\npb#1{{Nucl. Phys.} {\bf B#1}}
\def\mpl#1{{Mod. Phys. Lett} {\bf A#1}}
\def\ijmpa#1{{Int.J.Mod.Phys.}{\bf A#1}}
\def\rmp#1{{Rev. Mod. Phys.} {\bf 68#1}}


\end{document}